\documentclass[preprint2,numberedappendix]{emulateapj-rtx4}
\usepackage{graphicx,natbib}
\usepackage{bm,url,color}
\usepackage{amsmath}
\usepackage{amssymb}
\usepackage{graphics}
\usepackage{euscript}

\def\Sh{\mbox{{\rm Sh}}}
\def\tiSh{\widetilde{\rm S}{\rm h}}
\def\kf{k_{\rm f}}
\def\Pm{\mbox{\rm Pr}_M}
\def\Rm{\mbox{Re}_M}

\def\Rey{\mbox{\rm Re}}

\def\urms{u_{\rm rms}}
\def\tildeurms{\widetilde{u}_{\rm rms}}

\newcommand{\EQ}{\begin{equation}}
\newcommand{\EN}{\end{equation}}
\newcommand{\EQA}{\begin{eqnarray}}
\newcommand{\ENA}{\end{eqnarray}}

\newcommand{\EEq}[1]{Equation~(\ref{#1})}
\newcommand{\EEqs}[2]{Equations~(\ref{#1}) and~(\ref{#2})}

\newcommand{\Fig}[1]{Fig.~\ref{#1}}
\newcommand{\FFig}[1]{Figure~\ref{#1}}

{}
{}

{}
{}
{}
{}
{}
{}
{}
{}
{}
{}
{}
{}
{}
{}

{}
{}
{}
{}

\newcommand{\xxx}{\hat{\mbox{\boldmath $x$}} {}}
\newcommand{\yyy}{\hat{\mbox{\boldmath $y$}} {}}

\newcommand{\UU}{\mbox{\boldmath $U$} {}}

\newcommand{\BB}{\mbox{\boldmath $B$} {}}

\newcommand{\JJ}{\mbox{\boldmath $J$} {}}

\newcommand{\AAA}{\mbox{\boldmath $A$} {}}

\newcommand{\ff}{\mbox{\boldmath $f$} {}}

\newcommand{\FF}{\mbox{\boldmath $F$} {}}

\newcommand{\nab}{\mbox{\boldmath $\nabla$} {}}

\newcommand{\SSSS}{\mbox{\boldmath ${\sf S}$} {}}

\newcommand{\DDD}{{\cal D} {}}

\def\Pm{{\rm Pr}_{\rm M}}
\def\Rm{{\rm Re}_{\rm M}}

\def\half{{\textstyle{1\over2}}}

\def\onethird{{\textstyle{1\over3}}}

\newcommand{\yapj}[3]{ #1, {Astrophys. J.,} {#2}, #3}

\newcommand{\yapjl}[3]{ #1, {ApJ,} {#2}, #3}

\newcommand{\yana}[3]{ #1, {A\&A,} {#2}, #3}

\newcommand{\ygafd}[3]{ #1, {Geophys.\ Astrophys.\ Fluid Dyn.,} {#2}, #3}

\newcommand{\yjfm}[3]{ #1, {J.\ Fluid Mech.,} {#2}, #3}

\newcommand{\ypf}[3]{ #1, {Phys.\ Fluids,} {#2}, #3}

\newcommand{\yjetp}[3]{ #1, {Sov.\ Phys.\ JETP,} {#2}, #3}

\newcommand{\yprl}[3]{ #1, {Phys.\ Rev.\ Lett.,} {#2}, #3}

\newcommand{\ymn}[3]{ #1, {MNRAS,} {#2}, #3}

\newcommand{\ynat}[3]{ #1, {Nature,} {#2}, #3}

\newcommand{\ypre}[3]{ #1, {Phys.\ Rev.\ E,} {#2}, #3}
\newcommand{\yjpp}[3]{ #1, {J.\ Plasma\ Phys.,} {#2}, #3}

\newcommand{\yspu}[3]{ #1, {Sov.\ Phys.\ Usp.,} {#2}, #3}
\newcommand{\yps}[3]{ #1, {Phys.\ Scr.,} {#2}, #3}
\newcommand{\ynjp}[3]{ #1, {New\ J.\ Phys.,} {#2}, #3}
\newcommand{\yar}[3]{ #1, {Astron.\ Rep.,} {#2}, #3}

\newcommand{\yjour}[4]{ #1, {#2}, {#3}, #4}

\begin{document}

\title{Enhancement of small-scale turbulent dynamo by large-scale shear}
\medskip

\author{Nishant K. Singh$^{1,2}$}
\email{singh@mps.mpg.de}
\author{Igor Rogachevskii$^{3,2}$}
\email{gary@bgu.ac.il}
\author{Axel Brandenburg$^{2,4,5,6}$}
\email{brandenb@nordita.org}

\affiliation{
$^1$Max-Planck-Institut f\"ur Sonnensystemforschung,
Justus-von-Liebig-Weg 3, D-37077 G\"ottingen, Germany \\
$^2$Nordita, KTH Royal Institute of Technology
and Stockholm Univ., Roslagstullsbacken 23,
10691 Stockholm, Sweden \\
$^3$Department of Mechanical Engineering,
Ben-Gurion Univ. of the Negev, P. O. Box 653, Beer-Sheva
84105, Israel\\
$^4$JILA and Department of Astrophysical and Planetary Sciences,
Univ. of Colorado, Boulder, CO 80303, USA\\
$^5$Department of Astronomy, AlbaNova University Center,
Stockholm Univ., 10691 Stockholm, Sweden\\
$^6$Laboratory for Atmospheric and Space Physics, Univ. of Colorado,
Boulder, CO 80303, USA
}

\date{\today,~ $ $Revision: 1.239 $ $}

\begin{abstract}
Small-scale dynamos are ubiquitous in a broad range of turbulent flows
with large-scale shear,
ranging from solar and galactic magnetism to accretion disks,
cosmology and structure formation.
Using high-resolution direct numerical simulations we show that
in non-helically forced turbulence with zero mean magnetic field,
large-scale shear supports small-scale dynamo action, i.e., the
dynamo growth rate increases with shear
and shear enhances or even produces turbulence, which, in turn, further
increases the dynamo growth rate.
When the production rates of turbulent kinetic energy due to shear and forcing
are comparable, we find scalings for the growth rate $\gamma$
of the small-scale dynamo and the turbulent velocity $\urms$ with
shear rate $S$ that are independent of the magnetic Prandtl number:
$\gamma \propto |S|$ and $\urms \propto |S|^{2/3}$.
For large fluid and magnetic Reynolds numbers, $\gamma$, normalized by
its shear-free value, depends only on shear.
Having compensated for shear-induced effects on turbulent velocity,
we find that the normalized growth rate of the small-scale dynamo
exhibits the scaling,
$\widetilde{\gamma}\propto |S|^{2/3}$, arising solely from the
induction equation for a given velocity field.
\end{abstract}


\maketitle

\section{Introduction}
\label{Introduction}

In an electrically conducting turbulent fluid, the dynamo is a
fundamental phenomenon that can explain the origin of magnetic
fields in solar like stars, galaxies, accretion discs, etc.
Two types of turbulent dynamos are usually discussed in the literature:
large-scale and small-scale dynamos
\citep[see, e.g.,][]{M78,ZRS90,BS05}.
Magnetic field generation on scales smaller and larger than the
integral scale of turbulence are described as small-scale dynamo
(SSD) and large-scale dynamo (LSD), respectively.
The small-scale dynamos are ubiquitous and naturally find applications
in a broad range of topics such as
galactic magnetism \citep{KA92,RT16},
solar coronal heating \citep{ALA15},
accretion disks \citep{BN15},
cosmology and structure formation \citep{PMS14},
Riemannian manifolds \citep{SR13},
formation of the first stars in the Universe \citep{Schleicher+10}, etc.

The nature of the SSD depends strongly on the magnetic Prandtl number,
$\Pm= \nu/\eta$ \citep[see, e.g.,][]{KA92,SCT04,HBD04,SHB05,Bra11},
where $\eta$ is the magnetic diffusivity due to the
electrical conductivity of the plasma
and $\nu$ is its kinematic viscosity.
Random stretching of the magnetic field by smooth
velocity fluctuations in the viscous subrange of scales
describes the small-scale dynamo for $\Pm\gg1$
\citep[see, e.g.,][]{ZRS90,KR94,S98,KRS02,BS14,SCT04,HBD04,SHB05}.
The small-scale dynamo at low $\Pm$ is excited by the
turbulent inertial-range velocity fluctuations
(the spatially rough velocity field).
The growth rate of the small-scale dynamo
at low $\Pm$ is determined by the resistive magnetic diffusion scale
\citep[see, e.g.,][]{K68,VZ72,RK97,BC04,ISC07,SIC07,SSF12}.

Large-scale velocity shear is a common feature
of many astrophysical flows in, e.g., solar and stellar convective zones,
galaxies, accretion disks \citep[see, e.g.,][]{M78,ZRS90,BS05}.
In recent years, a non-helical
turbulent shear dynamo has been
discussed, where the presence of large-scale shear
in turbulence with zero mean kinetic helicity
yields a large-scale dynamo \citep[see, e.g.,][]{VB97,SOK97,
RK03,B05, KR08,YHS08,BRRK08,KKB08,SS09,SS10,SS11,SS14}.
The main conclusion from these studies is that a
combination of homogeneous non-helical turbulence and
large-scale shear is able to generate a large-scale magnetic
field without any mean kinetic helicity.
Like many other large-scale turbulent dynamos, they yield pronounced
large-scale magnetic structures.
Large-scale shear in non-helical turbulence also causes a
``vorticity dynamo'',
i.e., the excitation of a large-scale instability, resulting in
an exponential growth of the mean vorticity
\citep{EKR03,YHS08,KMB09}.

In turbulence with large-scale shear, the SSD
can be strongly affected by shear, notably
because turbulence itself can be produced by the shear.
However, the details related to the effect of shear on
the small-scale dynamo are unclear.
Recent analytical study by \cite{KLS11} has demonstrated that, for a given
random smooth velocity field, large-scale shear can support a
small-scale dynamo, such that the dynamo growth rate, which we
denote as $\widetilde{\gamma}$ arising solely from the induction
equation, increases with shear rate $S$
as $\widetilde{\gamma} \propto |S|^{2/3}$.
This is compatible with an upper bound for growth rates discussed
in \cite{Pro12}.

In this Letter, we study
the effects of large-scale shear on a small-scale dynamo using
high-resolution direct numerical simulations (DNS)
for different magnetic Prandtl numbers ranging from 0.5 to 10.
Using the budget equation for turbulent kinetic energy, we develop
a framework for identifying a
scaling function that we then determine in DNS.
In agreement with earlier work by \cite{KLS11},
we demonstrate that the growth rate of a SSD
in non-helically forced turbulence increases with the shear rate.

\section{Budget equation}
\label{Budget}

Since large-scale shear can affect the turbulent velocity,
we start with a theoretical analysis based on the budget
equation for turbulent kinetic energy,
${\cal E}_{\rm K}=\half \overline{{\bm u}^2}$,
assuming incompressibility \citep{MY71}:
\begin{eqnarray}
{D{\cal E}_{\rm K} \over Dt} + {\rm div} \, {\bm \Phi}_{\rm K} &=&
-\overline{u_i \, u_j} \, \nabla_j
\overline{U}_i + \overline{{\bm u} {\bm \cdot} {\bm
f}_{\rm f}}  - \varepsilon ,
 \label{A1}
\end{eqnarray}
where
$D / Dt = \partial /\partial t + \overline{\bm U} {\bm \cdot}
{\bm \nabla}$ is the advective derivative, ${\bm u}$ is the fluctuating
velocity, $\overline{\bm U}$ is
the mean velocity, and $\varepsilon$
is the dissipation rate of ${\cal E}_{\rm K}$.
The term ${\bm \Phi}_{\rm K}= \rho^{-1} \overline{{\bm u} \, p}
+ \overline{{\bm u} \, u^2}/2$ includes
the third-order moments that determine the flux of ${\cal E}_{\rm K}$,
where $p$ are the pressure fluctuations, and
$\rho$ is the fluid density.
The term $\overline{{\bm u} {\bm \cdot}{\bm f}_{\rm f}}$
in \EEq{A1} describes
the production rate of turbulence caused by external forcing,
while the first term in the right hand side of \EEq{A1}
determines the turbulence production rate caused by
large-scale shear.

The Reynolds stresses in isotropic turbulence are \citep{MY71,EKRZ02}:
\begin{eqnarray}
\overline{u_i \, u_j} = {\overline{u^2} \over 3} \, \delta_{ij}
- {\nu_{_{\rm T}} \over 2} \, \left(\nabla_i \overline{U}_j + \nabla_j \overline{U}_i\right),
 \label{A2}
\end{eqnarray}
where $\nu_{_{\rm T}}$ is the turbulent viscosity
and $\delta_{ij}$ is the Kronecker tensor.
Let us consider for simplicity turbulence with a linear velocity shear,
$\overline{\bm U}=(0, Sx, 0)$, which results in anisotropy of turbulence.
However, the modification of the Reynolds stresses by
anisotropic turbulence does not change the turbulence production rate caused by
linear velocity shear; see Equations~(A33) and~(12) in \cite{EKRZ02}.
The dissipation rate of ${\cal E}_{\rm K}$ for large fluid Reynolds numbers is
estimated as $\varepsilon \sim {\cal E}_{\rm K}/\tau_{\rm f} =
u_{\rm rms}^3/2\ell_{\rm f}$ \citep{MY71},
while the turbulent viscosity
is estimated as $\nu_{_{\rm T}} \sim \ell_{\rm f} \, u_{\rm rms}/3$,
where $\ell_{\rm f}$ is the integral scale of the turbulence,
$u_{\rm rms} =\sqrt{\overline{{\bm u}^2}}$ and
$\tau_{\rm f}=\ell_{\rm f} / u_{\rm rms}$ is the characteristic
turbulent time based on the integral scale.

Substituting \EEq{A2} into \EEq{A1} we obtain:
\begin{eqnarray}
{D{\cal E}_{\rm K} \over Dt} + {\rm div} \, {\bm \Phi}_{\rm K} =
\overline{{\bm u} {\bm \cdot} {\bm f}_{\rm f}}
+ {\nu_{_{\rm T}} \over 2} \, S^2
- {u_{\rm rms}^3 \over 2\ell_{\rm f}} .
 \label{A4}
\end{eqnarray}
Depending on the value of shear, \EEq{A4} implies the following
scalings for $u_{\rm rms}(S)$ in stationary homogeneous turbulence:

(i) {\it Small shear}: the turbulent production rate caused by
the forcing is much larger than that caused by the shear, so that
$u_{0}^3 / 2\ell_{\rm f} \sim \overline{{\bm u}{\bm \cdot} {\bm f}_{\rm f}}$,
and $\nu_{_{\rm T}}=\nu_{_{\rm T}}^{(0)} \sim \ell_{\rm f} \, u_{0}/3$,
where $u_0=u_{\rm rms}(S=0)$.
For small shear, $u_{\rm rms}$ is weakly dependent on shear.

(ii) {\it Intermediate shear}: the turbulent production rates caused by
the forcing and the shear are of the same order, and
the balance, $u_{\rm rms}^3 / \ell_{\rm f} \sim \nu_{_{\rm T}}^{(0)} S^2$,
yields the following scaling: $u_{\rm rms} \propto |S|^{2/3}$.

(iii) {\it Strong shear}:  the turbulent production rate in
\EEq{A4} caused by the forcing can be neglected,
so that turbulence is produced only by shear.
The steady-state solution of the equation, $\nu_{_{\rm T}} S^2
- u_{\rm rms}^3/ \ell_{\rm f}=0$, yields the scaling $u_{\rm rms}
= S \, \ell_{\rm f}$.
This implies that for shear-produced turbulence, the {\em small-scale}
shear rate $u_{\rm rms} / \ell_{\rm f}$ cannot be much
smaller than the large-scale shear.

\section{Growth rate of the small-scale dynamo}
\label{Growth rate}

Let us consider first the case $\Pm \ll 1$,
when the resistive magnetic diffusion scale is much larger
than the Kolmogorov viscous scale.
This implies that the resistive scale is located inside the inertial
range of the turbulence, where the fluid motions are spatially rough.
The small-scale dynamo occurs due to random stretching of the magnetic
field by the turbulent velocity, while scale-dependent turbulent
magnetic diffusivity causes dissipation of the magnetic field.
At the resistive scale, the scale-dependent turbulent magnetic diffusivity
approaches $\eta$.
The strongest magnetic field stretching is at small scales,
i.e., at the resistive scale.
Therefore, the growth rate of the small-scale dynamo
(far from the threshold) in turbulence without
large-scale shear for $\Pm \ll 1$
is estimated as the inverse resistive time
\citep[see, e.g.,][]{K68,SIC07}:
\begin{eqnarray}
\gamma_0 \sim u_\eta / \ell_\eta \sim \tau_{\rm f}^{-1}
\, \Rm^{1/2},
\label{C1}
\end{eqnarray}
where $u_\eta= (\varepsilon \, \ell_\eta)^{1/3}$ is the
characteristic turbulent velocity at the resistive scale, $\ell_\eta$,
and $\Rm$ is the magnetic Reynolds number.

\begin{figure*}
\begin{center}
\includegraphics{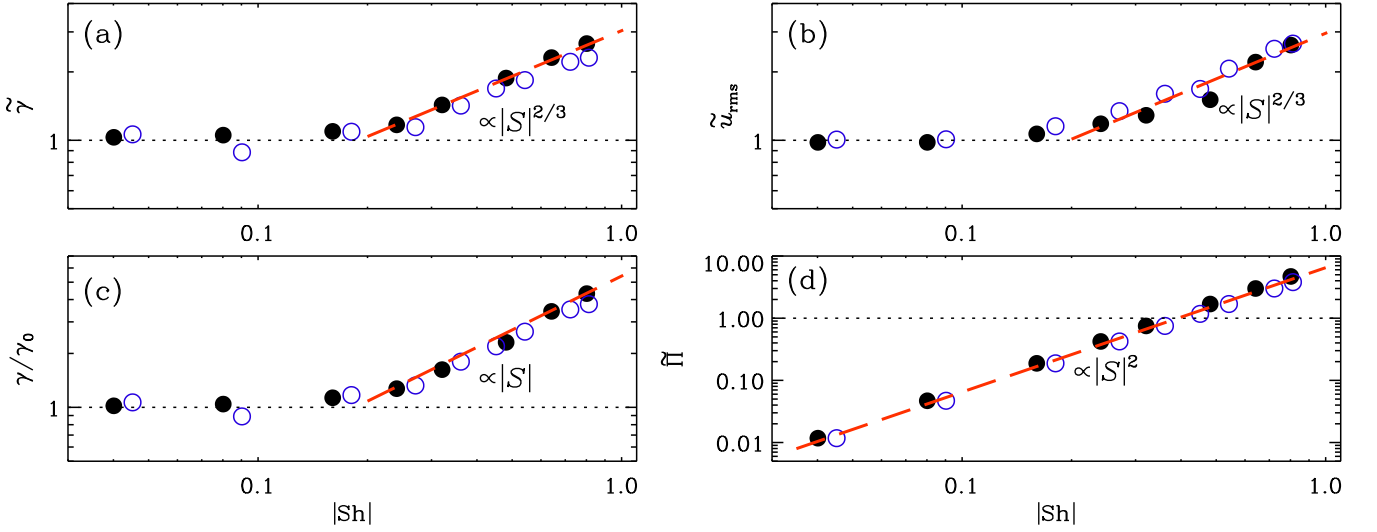}
\end{center}
\caption[]{
Shear dependencies of (a) $\widetilde{\gamma}$,
(b) $\widetilde{u}_{\rm rms}$, (c) $\gamma/\gamma_0$, and
(d) $\widetilde{\Pi}$, shown for two choices of $\Pm$, both sets
having $u_{\max}/c_{\rm s}=0.07$, and $|S|_{\max}/c_{\rm s}\kf=0.02$.
Filled (black) circles: $\Pm=0.5$ with $\gamma_0/u_0\kf=0.02$,
and $\Rm^{(0)}\equiv u_0 / \kf \eta=121$;
open (blue) circles: $\Pm=3$, with $\gamma_0/u_0\kf=0.05$,
and $\Rm^{(0)}\equiv u_0 / \kf \eta=148$.
}\label{gr_s_allPm}
\end{figure*}

Using \EEq{C1}, we assume that the growth rate of the small-scale
dynamo instability with large-scale shear, normalized by that without
shear, can be estimated as
\begin{eqnarray}
\frac{\gamma(S)}{\gamma_0} \sim \left(\frac{u_{\rm rms}(S)}{u_{0}} \right)^{1/2} \, \widetilde{\gamma}(S),
\label{C3}
\end{eqnarray}
where $\gamma_0=\gamma(S=0)$ and $u_0=u_{\rm rms}(S=0)$ represent the
dynamo growth rate and the rms velocity for $S=0$,
and $\widetilde{\gamma}(S=0)=1$.
Let us define the normalized rms velocity, $\tildeurms=u_{\rm rms}(S)/u_{0}$.
Here we assumed that the turbulent forcing scale,
$\ell_{\rm f}$, is independent of large-scale shear.
The contribution $\tildeurms^{1/2}$,
to the dynamo growth rate is caused by the effect of large-scale shear
on the turbulent velocity field, while
the function $\widetilde{\gamma}(S)$ determines the effect of
large-scale shear on the growth of the small-scale dynamo instability
for a given turbulent velocity field.
Thus the normalized growth rate, $\widetilde{\gamma}$,
which can be interpreted as a contribution
to the dynamo growth rate arising solely from the induction equation for a
given velocity field, can be expressed as
\begin{eqnarray}
\widetilde{\gamma}=\gamma(S) \Big/
\left(\gamma_0 \, \, \sqrt{\widetilde{u}_{\rm rms}} \right)\,.
\label{D5}
\end{eqnarray}
It is useful to define the ratio of the turbulent production rates
caused by shear and forcing,
\begin{eqnarray}
&& \widetilde{\Pi}=\frac{\Pi_S}{\Pi_{\rm f}}=
\frac{\nu_{_{\rm T}} S^2}{\urms f}
=\frac{2\pi S^2}{3\kf f} \, .
\label{D6}
\end{eqnarray}

For $\Pm \gg 1$, magnetic fluctuations are determined by
the smooth velocity field in the viscous subrange.
We assume that in turbulence with large-scale shear,
\EEq{C3} is also valid for $\Pm \geq 1$.
In the next section we perform DNS
to determine the scaling laws for $\gamma(S)/\gamma_0$
and $u_{\rm rms}(S)/u_{0}$.

\section{Numerical setup}
\label{DNS-setup}

We consider low-Mach-number compressible isothermal
magnetohydrodynamic turbulence with background
shear, $\overline{\bm{U}}^S=(0,Sx,0)$ with $S<0$, and a
white-noise nonhelical random
statistically homogeneous isotropic body force $\ff$
as the source of turbulent motions. The departure $\UU$
from the mean shear flow obeys
\begin{eqnarray}
&& {\DDD\UU\over\DDD t}=-\UU\cdot\nab\UU+SU_x\yyy
-c_{\rm s}^2\nab\ln\rho
+ \rho^{-1} \JJ\times\BB
\nonumber\\
&& \quad \quad \quad +\ff+\FF_{\rm visc},
\label{D1}\\
&&{\DDD\ln\rho\over\DDD t}=-\UU\cdot\nab\ln\rho-\nab\cdot\UU,
\label{D2}\\
&&{\DDD\AAA\over\DDD t}=-SA_y\xxx+\UU\times\BB-\eta\JJ \, ,
\label{D3}
\end{eqnarray}
where $\DDD/\DDD t\equiv \partial_t +Sx \,\nabla_y$
is the advective derivative
with respect to $\overline{\bm{U}}^S$,
$\BB=\nab\times\AAA$ is the magnetic field
in terms of the vector potential $\AAA$,
$\FF_{\rm visc}=\rho^{-1} \nab\cdot(2\nu\rho\SSSS)$
is the viscous force,
${\sf S}_{ij}=\half(\nabla_i U_j+\nabla_j U_i)-\onethird\delta_{ij}\nab\cdot\UU$
is the traceless rate of strain tensor,
$\JJ=\nab\times\BB/\mu_0$ is the current density,
$\mu_0$ is the vacuum permeability,
and $c_{\rm s}$ is the isothermal sound speed.
These equations are solved with shearing-periodic
boundary conditions using the {\sc Pencil Code}
(https://github.com/pencil-code).
It uses sixth-order explicit finite differences in space
and a third-order accurate time-stepping method.

We solve Equations~(\ref{D1})--(\ref{D3}) in a cubic domain
of size $L^3$ using $512^3$ or $1024^3$ spatial resolution
and choose $\kf=2.2 \, k_1$, where $k_1=2\pi/L$.
Thus the chosen stochastic forcing injects energy at scales close
to the box-scale.
This allows us to study the SSD in the absence of
the large-scale dynamo (LSD), or the so-called shear dynamo, that
is also expected to be excited in such a setup. However, the LSD
would require a reasonably larger scale separation.
Nevertheless, the growth rates measured from an early kinematic stage
predominantly reflect the growth of SSD which grows at a rate much
faster than that of the possible LSD.

The system of equations is characterized by the following set
of non-dimensional numbers:
\begin{eqnarray}
&&\Rm=\frac{\urms}{\eta \kf},\quad
\Rey=\frac{\urms}{\nu \kf},\quad
\Pm=\frac{\nu}{\eta}, \nonumber \\
&&\Sh=\frac{S}{u_0 \, \kf},\quad
\tiSh=\frac{S}{\urms \, \kf},\quad
\Sigma_{\rm f}=
\frac{c_{\rm s} \, \nu k_1^2}{f}
\label{control_pars}
\end{eqnarray}
where $\Rey$ and $\Rm$ are the fluid and magnetic Reynolds numbers,
$\Pm$ is the magnetic Prandtl number, $\Sh$ and $\tiSh$ are the
shear parameters based on $u_0$ and $\urms$, respectively,
and $\Sigma_{\rm f}$ characterizes the inverse forcing.

\section{DNS results}
\label{DNS-results}

In this section we discuss the results of DNS and
compare with the theoretical predictions.

\subsection{The dynamo growth rate and production of turbulence}

First we determine the small-scale dynamo growth rate
as a function of $S$.
To this end we drop the Lorentz force
in the momentum equation~(\ref{D1}).
In Fig.~\ref{gr_s_allPm} we show
the shear dependencies of (i) the normalized dynamo growth rate,
$\widetilde{\gamma}(S)$ which is defined by \EEq{D5},
(ii) the normalized rms velocity, $\tildeurms=u_{\rm rms}(S)/u_{0}$,
(iii) total growth rate $\gamma$, and
(iv) the ratio of turbulence production rates $\widetilde{\Pi}$,
for two values of $\Pm$ (smaller and larger than unity).
Figure~\ref{gr_s_allPm} demonstrates the existence of
the following scalings for intermediate shear
when the ratio $\widetilde{\Pi}$ is of the order of unity:
\begin{eqnarray}
\gamma/\gamma_0 \propto |S| , \quad \tildeurms \propto |S|^{2/3}
, \quad \widetilde{\gamma} \propto |S|^{2/3} \, .
\label{D7}
\end{eqnarray}
These scalings are independent of $\Pm$.
The small-scale dynamo growth rate increases with shear, which implies that
large-scale shear supports the small-scale dynamo.
The obtained DNS scaling for $\widetilde{\gamma}$ coincides with that
found by \cite{KLS11} from solution of the equation for the pair
correlation function of the magnetic field.
This equation was derived from the induction equation for a given
random smooth velocity field.

\begin{figure}
\begin{center}
\includegraphics[width=\columnwidth]{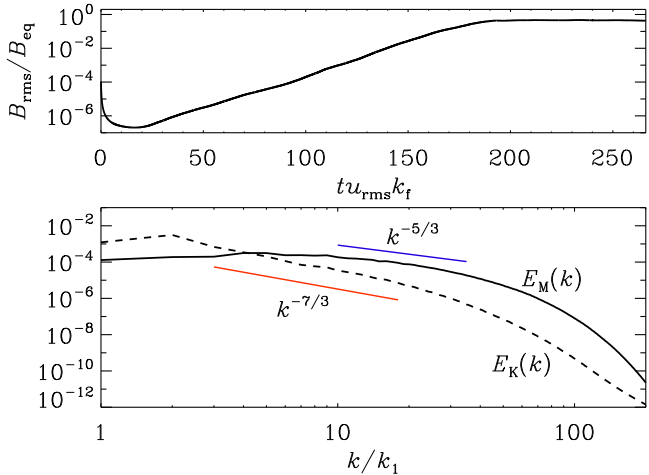}
\end{center}
\caption[]{Nonlinear evolution of $B_{\rm rms}/B_{\rm eq}$ (upper panel)
and spectra of magnetic $E_{\rm M}(k)$ (solid) and kinetic
$E_{\rm K}(k)$ (dashed) energies in the saturation stage
(lower panel) at $\Pm=10$ with $|\tiSh|=0.23$,
$u_{\rm rms}/c_{\rm s}=0.04$ and $|S|/c_{\rm s}\kf=0.009$.
}
\label{power_pm10}
\end{figure}

\subsection{Nonlinear stage of the SSD}

In Fig.~\ref{power_pm10} we plot the nonlinear evolution of
$B_{\rm rms}/B_{\rm eq}$ and spectra of magnetic, $E_{\rm M}(k)$,
and kinetic, $E_{\rm K}(k)$, energies
in the saturation stage,
where $B_{\rm eq}= (\mu_0\rho)^{1/2} \urms$ is the equipartition magnetic field.
Magnetic fluctuations reach saturation at the equipartition level
and have a short inertial range compatible with a $k^{-5/3}$ spectrum.
At larger scales, the velocity is compatible with a $k^{-7/3}$ spectrum which
is expected for anisotropic sheared fluctuations produced by tangling of
the large-scale gradient of the mean velocity by the background random
velocity field.
This spectrum was predicted analytically by \cite{L67}, detected
in atmospheric turbulence by \cite{WC72} and
confirmed in DNS by \cite{IY02}.

\begin{figure}
\begin{center}
\includegraphics[width=\columnwidth]{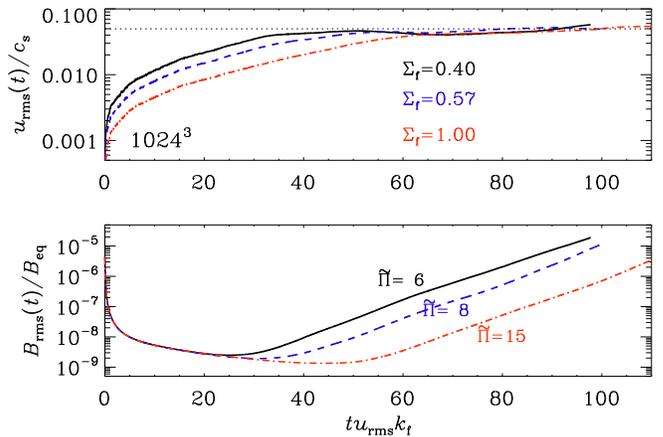}
\end{center}
\caption[]{
Temporal evolution of $\urms$ and $B_{\rm rms}$ from three runs
with solid, dashed, and dash-dotted curves in the order of
decreased forcing,
while all other parameters are the same:
$\Pm=10$, $\Rey=220$, $|\tiSh|=0.3$, $\urms/c_{\rm s}=0.05$, and
$|S|/c_{\rm s} k_{\rm f}=0.02$.
For all three cases we find $\gamma/\urms k_{\rm f}=0.146$.
}
\label{ts_1024}
\end{figure}

\FFig{ts_1024} shows results based on $1024^3$ simulations
at $\Pm=10$ for varying forcing strengths while keeping the shear-rate
$S$ as fixed.
The turbulence is produced by shear in all three cases, which yields
the same $u_{\rm rms}$ in the saturated state, thus resulting
in the same value for the shear parameter $\tiSh$.
The growth rates of
SSD are found to be identical, as would be expected from our above findings.
Note that the onset of the dynamo growth is delayed
for weaker forcing (e.g., red curve in Fig.~\ref{ts_1024}).

\subsection{Mean-flow generation}

\begin{figure}[h!]
\begin{center}
\includegraphics[width=\columnwidth]{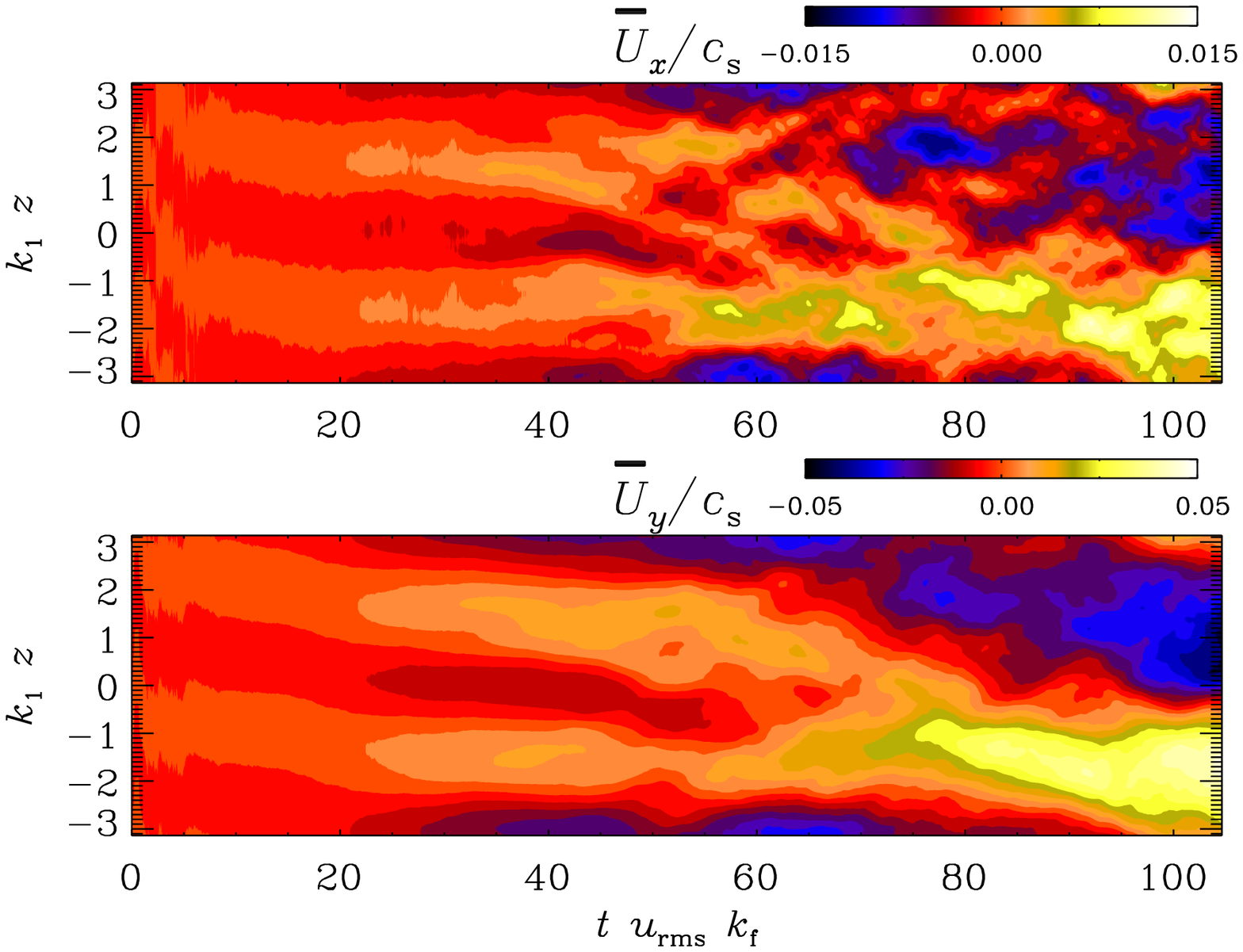}
\end{center}
\caption[]{
$\overline{U}_x$ (top) and $\overline{U}_y$ (bottom) as functions
of time and $z$ from a $1024^3$ simulation with $\Pm=10$,
$|\tiSh|=0.3$, $\widetilde{\Pi}=15$ and $u_{\rm rms}/c_{\rm s}=0.046$.
}
\label{flow_1024}
\end{figure}

\FFig{power_pm10} demonstrates the fact that shear fundamentally
modifies the nature of background turbulence, resulting in a $k^{-7/3}$
spectrum, which leads to the generation of a large-scale flow.
In \Fig{flow_1024} we show a space-time diagram of the mean flow
components, $\overline{U}_x$ and $\overline{U}_y$, where the mean
is obtained by applying a planar (here $xy$) average. Both
$\overline{U}_x$ and $\overline{U}_y$ spontaneously develop a mean
pattern in the direction normal to the shear plane.
Such a generation of mean flow has been first explored by \cite{EKR03}
and numerically demonstrated by \citet{YHS08,KMB09}.
The mean flow pattern begins to develop after a few tens of eddy
turnover time, $(u_{\rm rms} k_{\rm f})^{-1}$.
We found that $\overline{U}_y$ is about four
times stronger compared to $\overline{U}_x$ and both are excited in phase,
which is in agreement with \cite{KMB09}.

\subsection{Saturation of the shear parameter}

Based on $512^3$ simulations at magnetic Prandtl number $\Pm=0.5$ and
$3$, we showed in \Fig{gr_s_allPm}b
that the $u_{\rm rms}$ increases with shear. The dimensionless shear
parameter, $\tiSh$, defined with respect to $u_{\rm rms}(S)$, is
thus expected to approach saturation at large shear rates in the
regime of shear-produced turbulence. Here we check the saturation of $\tiSh$
by performing a suite of lower resolution, $128^3$, simulations
at $\Pm=1$. For a fixed shear rate, we explore the shear-produced turbulence
regime (i.e., the regime with $\widetilde{\Pi}>1$)
by successively decreasing the forcing strength, i.e., by
increasing $\Sigma_{\rm f}$; see \EEqs{D6}{control_pars}
for definitions of $\widetilde{\Pi}$ and $\Sigma_{\rm f}$, respectively.

\begin{figure}[h!]
\begin{center}
\includegraphics[width=\columnwidth]{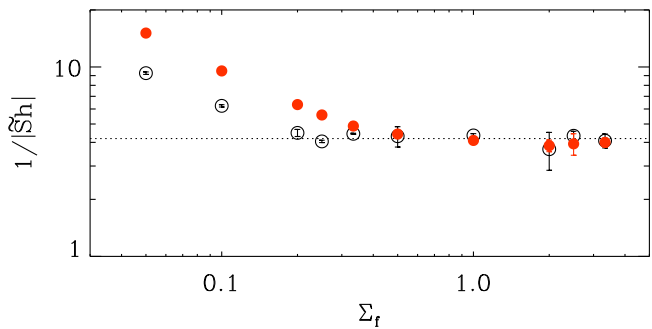}
\end{center}
\caption[]{
Saturation of the shear parameter $\tiSh$ as a function of $\Sigma_{\rm f}$
for two choices of shear-rate, $|S|/c_{\rm s} k_{\rm f}=0.022$
(black, open circles) and
$|S|/c_{\rm s} k_{\rm f}=0.013$ (red, filled circles), both at $\Pm=1$.
Larger values of $\Sigma_{\rm f}$
corresponds to the regime where the turbulence is predominantly produced by
shear. The dotted line represents $|\tiSh| = 3/4\pi$.
}
\label{u_saturation}
\end{figure}

In \Fig{u_saturation}, we demonstrate the saturation of the shear parameter,
$\tiSh$ as a function of $\Sigma_{\rm f}$. Two different choices of the
shear rate result in the overlap of $1/|\tiSh|$ at large values of
$\Sigma_{\rm f}$ and show saturation at a constant level corresponding to
$|\tiSh| \approx 3/4\pi$. Thus, in a realistic setup with subsonic turbulence,
such as the one being studied here, it would not be possible to explore values
of $|\tiSh|$ that are much larger than about $0.25$. Note that the abscissae
in \Fig{gr_s_allPm} correspond to $\Sh$ which is defined
with respect to $u_0$, instead of $u_{\rm rms}(S)$, and therefore
extend up to about unity.

The effect of shear on the SSD becomes noticeable only when shear-rate
exceeds certain threshold such that the turbulence production ratio,
$\widetilde{\Pi}$, becomes of order unity or larger.
This results in a narrow range of possible values for the shear
parameter to determine the scalings of the SSD growth rates versus shear
rate.
Thus, one interesting regime of a very strong shear
is not found in DNS.

\section{Discussion}
\label{Discussion}

It is worth noting that recent simulations by \cite{TC13}
of a prescribed deterministic non-helical flow with
large-scale shear and a superposition of small-scale
cellular deterministic flows, have shown that large-scale
shear reduces the small-scale dynamo growth rate.
In those simulations, the Navier-Stokes equation
is not used, so the effects of shear-produced turbulence,
whereby large-scale shear increases the turbulent velocity,
have been ignored.
As a result, shear suppresses the small-scale dynamo
by a sweeping effect, i.e., the shear de-correlates the small eddies
from the magnetic field by advecting the field.
On the other hand, in our study a different setup is used, where
large-scale shear fundamentally modifies and even produces turbulence,
and enhances the efficiency of the small-scale dynamo.
Notably, the generation of non-uniform mean velocity field, as shown
in \Fig{flow_1024}, was not observed in \cite{TC13}. This additionally
confirms the difference between their setup and ours.

Remarkably, we obtain from DNS the same scaling,
$\widetilde{\gamma}(S) \propto |S|^{2/3}$, for SSD growth rate
as was theoretically predicted by \cite{KLS11} for a given
velocity field.
Interestingly, the occurrences of intermittent shear bursts were found
to amplify the growth of the SSD in turbulent magnetoconvection
\citep{PBM13}.

\section{Conclusions}
\label{Conclusions}

Using DNS, it was demonstrated that
the small-scale dynamo growth rate increases with shear
in nonhelical turbulence.
The scalings for the growth rate of the small-scale dynamo,
$\gamma \propto |S|$, and for the turbulent velocity,
$\urms \propto |S|^{2/3}$, are independent of $\Pm$,
when the turbulent production rates caused by shear and forcing
are of the same order.
The contribution to the dynamo growth rate,
$\widetilde{\gamma}\propto|S|^{2/3}$, is also found to be independent of $\Pm$.
This contribution is determined solely by the
equation for the pair correlation function
of the magnetic field derived from the induction equation.

We found that large-scale shear has the following three different effects
that are relevant for turbulent small-scale dynamo:
\begin{enumerate}
\item [$\bullet$] Direct effect of shear on the generation of
small-scale magnetic fields through the induction equation.
\item [$\bullet$] Production of turbulence by the shear which
further enhances the SSD action.
\item [$\bullet$] Generation of large-scale non-uniform motions
due to interaction of turbulence with mean shear by the vorticity dynamo,
which in turn produces new large-scale shear, thus enhancing the
SSD.
\end{enumerate}

\acknowledgments
We have benefited from stimulating discussions with Pallavi Bhat,
Nathan Kleeorin, Igor Kolokolov, Dhrubaditya Mitra, Matthias Rheinhardt,
Kandaswamy Subramanian, and Steve Tobias.
IR thanks NORDITA for hospitality and support during his visits.
This work has been supported in parts by
the NSF Astronomy and Astrophysics Grants Program (grant 1615100),
the Swedish Research Council grant No.\ 621-2011-5076 and the
Research Council of Norway under the FRINATEK grant No.\ 231444.
We acknowledge the allocation of computing resources provided by the
Swedish National Allocations Committee at the Center for
Parallel Computers at the Royal Institute of Technology in Stockholm,
and by CSC -- IT Center for Science Ltd.\ in Espoo, Finland, which is
administered by the Finnish Ministry of Education (project 2000403).

\end{document}